# An interior solution of constant density to Gurses' metric

S.B.P. Wickramasuriya
Department of Mathematics, University of Kelaniya, Kelaniya, Sri Lanka

Abstract. We give here an interior metric which matches Gurses' metric. The metric given here is a sphere of electrically counterpoised dust (ECD) of constant density. We use Lane Emden equation for obtaining the interior solution.

Key words. Majundar Papapetrou metrics, Lane Emden equation, electrically counterpoised dust.

We present an interior solution composed of electrically counterpoised (extremely charged) dust which matches with the solution obtained by Gurses in 1998 [1]. This is also discussed by Varela [2]. The interior solution is obtained by considering a nonlinear second order differential equation which occurs in astrophysics. The author encountered this problem in 1972 [3]. We first obtain the equation for ECD.

Following applies to metrics of the form
$$ds^2 = e^{2U} dt^2 - e^{-2U}(dx^{1^2} + dx^{2^2} + dx^{3^2})$$
which we write as $ds^2 = e^{2U} dt^2 - e^{-2U} \gamma_{\alpha\beta} dx^\alpha dx^\beta$

For electrically counterpoised dust (ECD) distributions
$$\gamma^{\mu\nu}\left(e^{-U}\right)_{1\mu\nu} = -4\pi\rho e^{-3U} .$$
The vertical stroke "$_1$" denotes covariant differentiation with respect to the metric $d\sigma^2 = \gamma_{\alpha\beta} dx^\alpha dx^\beta$.

This can be written in spherical polar coordinates as
$$\frac{1}{r^2}\frac{d}{dr}\left(r^2 \frac{de^{-U}}{dr}\right) = 4\pi\rho \, e^{-3U} .$$
which is a nonlinear second order differential equation.
Putting $e^{-U} = ky, r = lx$ this can be transformed to the Lane Emden equation [4]

$$\frac{1}{x^2}\frac{d}{dx}\left(x^2 \frac{dy}{dx}\right) = -y^n \quad (n=3)$$

The solutions for which the initial conditions are $y(0)=1, y'(0)=0$, are known as Emden functions and have been tabulated. This equation occurs frequently in problems involving astrophysics.

$\left(e^{-U}\right)(0) = ky(0) = k, \ \left(e^{-U}\right)'(0) = 0$
where the prime denotes differentiation.

Let us consider a uniform sphere of density $\rho$ and radius $a$. For $r > a$ we have a vacuum with only the electromagnetic field being present. In vacuum we assume Gurses' solution.

$$e^{-U} = 1 - \frac{m_0}{a} + \frac{m_0}{r}$$

We match the interior and exterior solutions at $r = a$

$$(e^{-U})(a) = ky\left(\frac{a}{l}\right) = 1 \qquad (1)$$

The derivative should also be continuous

$$(e^{-U})'(a) = \frac{k}{l} y'\left(\frac{a}{l}\right) = -\frac{m_0}{a^2} \qquad (2)$$

Equations (1) and (2) give

$$y\left(\frac{a}{l}\right) + \frac{a}{m_0}\left(\frac{a}{l}\right) y'\left(\frac{a}{l}\right) = 0$$

which can be written as $y(x_0) + \left(\frac{a}{m_0}\right) x_0 y'(x_0) = 0, \quad x_0 = \frac{a}{l}$

For us to be able to match the two metrics at $r = a$ we have to find $x_0$ and $\frac{a}{m_0}$ such that they satisfy the above relation. That this can be done is intuitively clear on an inspection of the graph of $-\frac{(x_0 y'(x_0))}{y(x_0)}$ against $x_0$, for $0 < x_0 < 6.89685$, where 6.89685 is the first zero of $y(x)$.

We have constructed a source for Gurses' solution. The source is a sphere of constant density of ECD. It will be interesting to examine the maximum possible density for a given a.